\documentclass[]{spie}  

\usepackage{amsmath,amsfonts,amssymb}
\usepackage{graphicx}
\usepackage{tabularx}
\usepackage{caption} 
\usepackage[colorlinks=true, allcolors=blue]{hyperref}

\title{The GLINT nulling interferometer: improving nulls for high-contrast imaging}

\author[a]{Eckhart Spalding}
\author[b]{Elizabeth Arcadi}
\author[b]{Glen Douglass}
\author[b]{Simon Gross}
\author[c]{Olivier Guyon}
\author[d]{Marc-Antoine Martinod}
\author[a]{Barnaby Norris}
\author[a]{Stephanie Rossini-Bryson}
\author[a]{Adam Taras}
\author[a]{Peter Tuthill}
\author[c]{Kyohoon Ahn}
\author[c]{Vincent Deo}
\author[e]{Mona El Morsy}
\author[c]{Julien Lozi}
\author[c]{Sebastien Vievard}
\author[b]{Michael Withford}
\affil[a]{Sydney Institute for Astronomy, School of Physics, University of Sydney, Sydney NSW 2006, Australia}
\affil[b]{MQ Photonics Research Centre, Macquarie University, Sydney NSW 2109, Australia}
\affil[c]{National Astronomical Observatory of Japan, Subaru Telescope, National Institutes of
Natural Sciences, Hilo, HI 96720, USA}
\affil[d]{Institute of Astronomy, KU Leuven, Celestijnenlaan 200D, 3001 Leuven, Belgium}
\affil[e]{Department of Physics \& Astronomy, University of Texas, San Antonio, San Antonio, TX 78249, USA}

\authorinfo{Further author information: (Send correspondence to E.S.)\\E-mail: eckhart.spalding@sydney.edu.au}

\begin{document} 
\maketitle

\begin{abstract}
GLINT is a nulling interferometer downstream of the SCExAO extreme-adaptive-optics system at the Subaru Telescope (Hawaii, USA), and is a pathfinder instrument for high-contrast imaging of circumstellar environments with photonic technologies. GLINT is effectively a testbed for more stable, compact, and modular instruments for the era of $\sim$30m-class telescopes. GLINT is now undergoing an upgrade with a new photonic chip for more achromatic nulls, and for phase information to enable fringe tracking. Here we provide an overview of the motivations for the GLINT project and report on the design of the new chip, the on-site installation, and current status.
\end{abstract}


\section{INTRODUCTION}
\label{sec:intro} 

Optical-infrared (OIR) astronomy stands on the threshold of the era of Extremely Large Telescopes (ELTs), which will mark the expansion of ground-based primary collecting apertures to $\approx$2.5-5$\times$ that of the largest OIR telescopes of today. As we anticipate the commensurate increases in light-gathering power and achievable resolution, early planning is also taking place for future flagship space missions, which will have the goal of finding and characterizing habitable worlds in our stellar neighborhood.  

To take full advantage of future platforms and narrow in on some of their most compelling science cases, new techniques will be necessary to recover faint signals at ever finer angular resolutions, and to do so as part of instruments installed in ever larger and more complicated facilities. The highest resolution information will necessarily have to be sieved out interferometrically--- coronagraph inner working angles are usually limited to $\approx 1-3 \lambda/D$ (e.g., \cite{guerri2011apodized,nguyen2022gpi}), but sparse-aperture masking or nulling interferometry can use single apertures (solid or multisegmented) to obtain resolutions as small as $\approx 0.5 \lambda/D$ (e.g., \cite{cheetham2016sparse,ray2023textit}). 

Upscaling the size and complexity of instruments to accommodate larger telescopes will open up unwanted additional channels of thermal noise, vibrations, flexure, and other systematics that will complicate aperture cophasing and wavefront control, as well as introducing various additional overheads. Fortuitously, methods from the telecommunications field are now seeping into astronomy, giving rise to astrophotonics: the application of miniaturized, rigid, and highly replicable optics to astronomy (e.g., \cite{bland2009astrophotonics,gross2015ultrafast,jovanovic20232023}). Astrophotonic instruments offer the possibility of decoupling the size of the instrument from that of the telescope, maintain a rigid phase matching between waveguides, reduce the need for cryogenics, can easily be modified in a modular fashion, and have manufacturing workflows which are highly repeatable.

The Guided-Light Interferometric Nulling Technology (GLINT) instrument is a pathfinder for maturing the technology of integrated and miniaturized photonics which replace traditional bulk optics in interferometric beam combination, and to mask out contaminating light from bright host stars to observe their circumstellar environments at tiny angles. Active since 2016 at Subaru Telescope on Mauna Kea (and briefly with a sister instrument at the AAT near Coonabarabran, Australia; see Table \ref{tab:versions_glint}), the core of the GLINT instrument is a glass chip containing a 3D arrangement of single-mode waveguides made with a laser inscription (ULI) which causes a $\sim$10$^{-2}$ change in the index of refraction within the glass to propagate single-mode wavefronts. The waveguides are designed to ``null'' stellar photons in the science channel by diverting on-axis light and transmitting the off- and on-axis light to separate outputs. GLINT is the first instrument to perform on-chip nulling.

In this proceeding we describe some of the potential science areas which set the context for this work, explain the functionality of GLINT, and describe the specific motivations for the ongoing upgrade, the project's current status, and possible future directions.

\section{Future science areas}

A host of science areas stand to benefit from the maturation of photonic nulling, both on 8-m class telescopes of today and 30-m class telescopes of tomorrow (e.g., \cite{defrere_review}). Among targets which can already be partially resolved with large telescopes are the dusty outflows of post-main-sequence stars or young stellar objects (YSOs) in star-forming regions, but their inner radii remain understudied \cite{blind2011incisive,persi2019infrared,crowe2024near}. Imaging the outflow morphologies in the OIR at $\sim$10s mas can help disentangle their physics by imaging close to the outflow formation zones. While some systems have been probed at those scales using sparse aperture masking\cite{lykou2015dissecting}, nulling would serve to decrease the photon noise contribution from the star.

\begin{figure} [ht]
\begin{center}
\begin{tabular}{c} 
\includegraphics[height=4.65cm, trim={4.5cm 1.2cm 4.5cm 1.2cm}, clip=True]{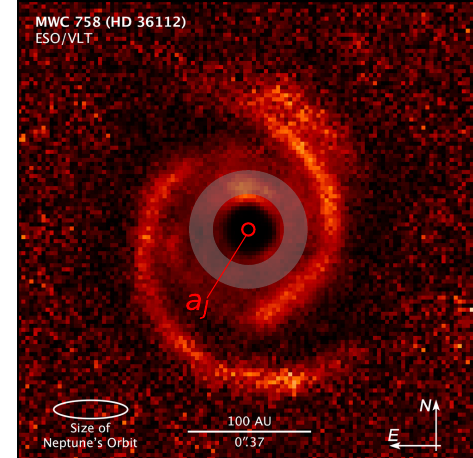}
\includegraphics[height=4.65cm, trim={3.5cm 6.5cm 15cm 6.6cm}, clip=True]{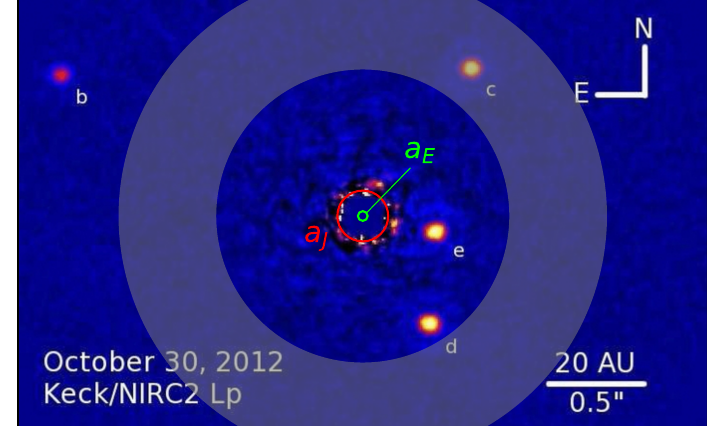}
\includegraphics[height=4.65cm, trim={17cm 3.7cm 2.6cm 4.4cm}, clip=True]{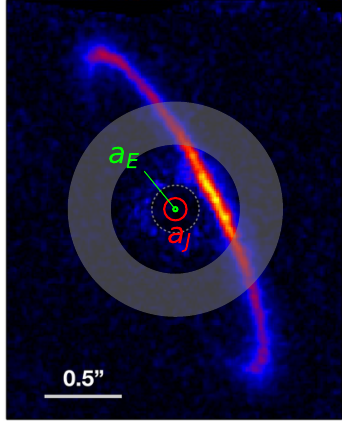}
\end{tabular}
\end{center}
\caption[example] 
{ \label{fig:scales} 
A few iconic directly imaged systems, with overplotted lines to illustrate the comparative scales of our solar system, much of which would be within the inner working angles of coronagraphs (masked inner regions in these plots). Resolving these innermost structures will require interferometry. Green circles represent the face-on orbit of Earth $a_{E}=1$ AU, red circles that of Jupiter $a_{J}=5.2$ AU, and grey shading the approximate extent of the Kuiper Belt, 30-50 AU. \textbf{Left:} the protoplanetary disk around MWC 758 in polarized $Y$-band (1.0 $\mu$m) from VLT/SPHERE. For a nulling resolution element $\lambda/2B$ in $H$-band, 3.2 such elements would fit across the diameter of Jupiter's orbit with a 8-m telescope, and 11.8 with a 30-m telescope. (Image from NASA and STScI, based on \cite{benisty2015asymmetric}; in public domain.) \textbf{Middle:} the exoplanets of the HR 8799 system from Keck/NIRC2 in $L'$-band (3.8 $\mu$m). A total of 12.3 and 46.1 $H$-band nulling resolution elements would fit within Jupiter's orbit with an 8-m and 30-m telescope. (Adapted from \cite{currie2014deep} Fig. 1; original © AAS. Reproduced with permission.) \textbf{Right:} the debris disk around HR 4796A from Gemini-S/GPI in polarized $K1$-band (2.0 $\mu$m). A total of 7.0 and 26.3 $H$-band nulling resolution elements would fit within Jupiter's orbit with a 8-m and 30-m telescope. (Adapted from \cite{perrin2015polarimetry} Fig. 7; original © AAS. Reproduced with permission.) }
\end{figure}

Circumstellar disk imaging in scattered light also requires overcoming high contrast at tiny angular scales, and while a veritable zoo of disks have been imaged with coronagraphic instruments at AO-equipped telescopes, only their outer regions have been uncovered in the OIR, again at inner working angles $>\lambda/D$. In the near infrared this frequently means imaging disk scales of several or $\sim$10s AU at best--- larger than our inner solar system \cite{garufi2024sphere,hughes2018debris,benisty_2023}. (See Fig. \ref{fig:scales}.)

Higher-resolution imaging of protoplanetary disks and slightly older transition disks will reveal planet-formation processes at the locations of the water ice line at $\sim$1s AU, which is thought to be critical for setting the locations of planet formation \cite{muller2021water}. (Structure at such orbits around stars in Upper Scorpius or Taurus-Auriga, the nearest large star-forming regions $\approx$120-160 pc away \cite{roccatagliata20203d}, requires imaging down to $\sim$mas.) Resolved spirals, eddies, or gaps on the scale of the inner solar system could provide hints of forming solar-system-analogue planets and even enable their indirect characterization \cite{dong2019observational}. Accreting planets may even be directly detected, and the (very small) handful of current candidates at $\sim$10s AU offer tantalizing hints of what else is to be found on smaller orbits \cite{sallum2015accreting,haffert2019two}.

In more evolved systems, cold debris disks of rocky material retain information about the dynamical history of the system. Though such disks can be on wide orbits ($\sim$10s to $\sim$100 AU), interferometric imaging has a role to play in systems further away, and of course to capture finer detail, and remove stellar photon noise with nulling. Exozodiacal disks, which are optically thin but are bright in integrated light, will be a potentially important noise source in future space-based exoplanet direct imaging to look for habitable worlds. Nulling has already unveiled part of the luminosity function of exozodiacal disks in the habitable zones around FGK-type stars $<$30 pc away \cite{weinberger2015target,ertel2020hosts}, but their morphologies (and consequently the danger of confusion noise) remain largely uncharacterized. There are essentially no observational constraints on the luminosity function around M-dwarfs, whose habitable zones are at angles $\sim$1s to $\sim$10s mas in the solar neighborhood, $<$30 pc (e.g., \cite{turnbull2012search,golovin2023fifth}). 

The field of exoplanets remains particularly photon-starved even though various detection techniques have uncovered thousands of extrasolar worlds. Often little more is known of such worlds than the most basic parameters, like the period of the planet's orbit, and the planet-to-star radius ratio or upper limits on the planet's mass. The number of planets which have been atmospherically characterized using spectra of significant resolution and signal-to-noise are in the $\sim$10s, and they tend to be giant planets with masses of Neptune or greater \cite{madhusudhan2019exoplanetary,chauvin2023direct}. As the exoplanet field transitions from primarily detecting other worlds to studying them in detail, direct or interferometric imaging will be critical to the task. Even without high resolution spectroscopy, imaging observations which are complimentary for a given system can potentially break important degeneracies (like imaging in combination with radial velocity measurements, which would break degeneracies in cooling models). 

In the most favorable cases, contrast ratios due to thermally emitted light of young, hot planets in the near- and thermal infrared are $\sim$$10^{-5}$ or $\sim$$10^{-6}$, and in exceptional cases $\sim$$10^{-4}$ (e.g., \cite{traub2010direct,bonnefoy2011high}). In reflected light at visible wavelengths the required contrasts are still more demanding, around $\sim$$10^{-8}$ to $\sim$$10^{-11}$, depending on the planet's orbital configuration (e.g., \cite{lovis2017atmospheric,carrion2021catalogue}). Imaging planets on solar-system-like orbits around stars further away than a few 10s pc with coronagraphs will become prohibitive, such that early target lists for the coronagraphic Habitable Worlds Observatory (HWO) include only systems within 50 pc, roughly a third of the distance to the nearest star-forming regions \cite{mamajek2024nasa,tuchow2024hpic}. In addition, a full 72 \% of known stars within 25 pc to Earth are M-dwarfs\cite{golovin2023fifth}, and drilling down into the physics of individual systems even in our immediate solar neighborhood will inevitably require observations at angles frequently within the inner working angles of coronagraphs.




\section{A nuller on a chip}

\subsection{The heritage of GLINT}

Photonic chips have been used for astronomical beam combination for more than two decades, but only in a small handful of instruments until very recently. The Infrared-Optical Telescope Array's IONIC instrument (2000-2006) on Mt. Hopkins in the U.S. was the first to successfully perform beam combination with chips with dimensions of a few cm, combining two and three telescope beams in $H$-band \cite{berger2001integrated,berger2003integrated,monnier2004first}. The VLTI/VINCI instrument (2001-2004) used a chip to combine $K$-band beams from two VLT telescopes \cite{lebouquin2006integrated}. VLTI/PIONIER (2010-) combines four VLTI beams, either from the UT or AT telescopes, across six baselines in $H$-band \cite{benisty2009integrated,le2011pionier}. VLTI/GRAVITY (upgraded to GRAVITY+) (2016-) combines the science $K$-band (2.2 $\mu$m) beams of all four UT telescopes with a cryogenically-cooled 5 cm $\times$ 2 cm chip \cite{jocou2014beam,perraut2018single}. All of these instruments featured chips based on layered silica and etching techniques, made in collaboration with CEA-Leti.

OIR nulling interferometers have also been few in number. The MMT/BLINC instrument (1998-2010) operated across 2 to 25 $\mu$m and used a 4 m baseline across the MMT and Magellan-Baade telescope apertures \cite{hinz2000blinc}. The Keck Interferometer Nuller (KIN) (2003-2012) operated in $N$-band across the 85 m baseline of the twin Keck telescopes \cite{colavita2010keck}. Currently-operating nullers include the Palomar Fiber Nuller (PFN) (2008-) at the 5.1-m Palomar Hale telescope, which combines two beams across a 3.4 m baseline in $Ks$-band (2.15 $\mu$m) by focusing them onto the opening of a single-mode fiber (SMF). When one beam has a $\pi$ relative phase shift, the field contributed by the on-axis star becomes antisymmetric and does not couple into the SMF \cite{haguenauer2006deep,serabyn2019nulling}. The Large Binocular Telescope Interferometer (LBTI) (2010-), the technical successor to BLINC and the scientific successor to the KIN, can combine 11 $\mu$m beams in nulling mode from 2$\times$8.4-m mirrors coaxially and with a relative $\pi$ phase shift by sending them in opposite directions through a 50-50 beamsplitter housed within a cryostat \cite{hinz2016}. The science beam (from off-axis light) proceeds to the detector, while the stellar (on-axis) light is diverted.

The Dragonfly project was a proof-of-concept demonstration of the use of a 3 cm $\times$ 2 cm glass chip to remap subapertures from the 3.9-m AAT telescope pupil through 3D waveguides with outputs into a linear fiber array, where the beams could be conveniently interfered off-chip, though not in nulling mode \cite{jovanovic2012starlight}. The astronomical $H$-band was a natural early band for photonics experimentation, as it overlaps with the telecommunication industry's heavily-used ``C-band,'' a transmissive window at $\approx$1.6 $\pm 0.2$ $\mu$m in optical fibers, where low losses are bounded in wavelength space by a spike of higher opacity at $\approx$1.4 $\mu$m due to OH impurity absorption, and multiphonon absorption at $\gtrsim$1.8 $\mu$m whereby photons are absorbed and their incident energies are converted into multiple vibrational modes in the material (e.g., \cite{cherin1983introduction}). GLINT builds on the basic design of Dragonfly by performing pupil remapping \textit{and} beam combination on a photonic chip, and takes advantage of the flexibility on-chip beam combination to enable nulling.

\subsection{Previous nulling with GLINT}

The primary criterion for a nulling observation is the null depth, defined as the ratio of the minimum and maximum transmissive fringes $I$: 

\begin{equation}
\mathcal{N} \equiv \frac{I_{min}}{I_{max}} = \frac{1-V}{1+V}
\label{eqn:null}
\end{equation}

\noindent
where $V$ is the fringe visibility, the quantity more often used in the case of non-nulling interferometry where the fringes are visible on the detector \cite{serabyn2021nulling}. The deeper the instrumental null, the more the host star light is masked out, but the null depth will be degraded by chromaticity across the bandpass. (In practice the starlight will also leak through the null due to phase noise or the extended disk of the star.)

\begin{table}[ht]
\caption{The evolution of GLINT} 
\label{tab:versions_glint}
\centering
\footnotesize
\begin{tabularx}{\textwidth}{|*{5}{X|}} 
\hline
 & \textbf{Subaru, 2016-2018} & \textbf{AAT, 2016-2017 (``GLINT-South'')} & \textbf{Subaru, 2018-2023} & \textbf{Subaru, 2024-} \\
\hline
\textbf{Number of apertures/baselines} & 2/1 & 2/1 & 4/6 & 3/3 \\
\hline
\textbf{Baseline lengths} & 5.55 m & 2.7 m & 2.15-6.45 m & 4-6 m \\
\hline
\textbf{$\lambda/(2B)$} & 29 mas & 59 mas & 25-74 mas & 27-40 mas \\
\hline
\textbf{Interference}$^{1}$ & DC & DC & DC & TRI \\
\hline
\textbf{AO} & AO188+SCExAO & Internal tip-tilt only & AO188+SCExAO & AO3k$^{3}$+SCExAO \\
\hline
\textbf{Detection} & Photoreceivers (photometry) & Photodiodes (photometry) & C-RED 2 (spectra) & C-RED 1 (spectra) \\
\hline
\textbf{Remarks} & Measured stellar diameters $\sim$19-31 mas  & Measured stellar diameters $\sim$39-54 mas without high-order AO; null precision $\sim$$10^{-3}$ (on-sky) & Measured stellar diameters $\sim$11-24 mas; instrumental $\mathcal{N}$$\lesssim$$10^{-3}$ (off-sky)$^{2}$ & Tricouplers will provide phase information; integrated achromatic phase shifters; MLA glued to chip \\
\hline
\textbf{Refs} & \cite{norris2020first} & \cite{lagadec2018glint,lagadec2021glint} & \cite{martinod2021scalable,norris2023machine} & (These proceedings) \\
\hline
\end{tabularx}
\begin{minipage}{\textwidth}
\vspace{0.1cm}
\small  1: DC: directional couplers; TRI: tricouplers \\
\small  2: Note null depth fluctuations and precision in the null depth are important for determining the astrophysical null \\
\small  3: anticipated
\end{minipage}
\end{table}

Up until now, the waveguides in GLINT chips have nulled with directional couplers, whereby two waveguides come close together and the wavefronts they carry interfere through the coupling of their evanescent fields which penetrate into the glass bulk. On-axis light is nulled by applying an optical path difference of $\pi/2$ in one of each pair of interfering waveguides, which directs on- and off-axis light into separate output channels. The null depth, fluctuations, and precision in the nulled channel output constrains the astrophysical contrast of off-axis light with the host star. Reduction methods such as the Numerical Self-calibration Method make use of the statistical distribution of the time-modulating signal to alleviate systematics introduced by phase noise and even obviate the need for calibrator stars \cite{hanot2011improving}. 

The pupil remapping of GLINT and Dragonfly chips kept pathlengths matched, to within constant offsets \cite{jovanovic2012starlight,jovanovic2012progress}. The chips also removed the imperative of non-redundancy among all interferometric baselines, because now wavefronts could be confined to custom waveguides and be interfered in whichever combinations desired. This can maximize throughput and Fourier coverage, and could in principle enable the carving up of the entire aperture into different baselines.

The ancillary layout of GLINT includes optics which focus and inject the light into the chip, a deformable mirror (DM) with discrete segments to enable phase offsets in the beams injected into the chip; and ``pigtail'' fiber outputs from the chip that lead the light to either a photoreceiver or to a dispersive element for spectra of $R\approx160$ \cite{norris2020first,martinod2021scalable}. 

Previous versions of GLINT achieved null depths as good as $\mathcal{N}$$\sim$$10^{-3}$ with $\sim$$10^{-4}$ precision (Table \ref{tab:versions_glint}). While encouraging, contrasts of $\sim$$10^{-6}$ need to be overcome to image massive young planets, which will require deep, precise, and achromatic nulls. There are a few routes for approaching this: waveguide designs which produces more achromatic nulls, the addition of closed-loop phase control based on the chip outputs, and the reduction of other noise sources from outside the chip. 

\subsection{Motivations for a new design}

\begin{figure} [ht]
\begin{center}
\begin{tabular}{c}
\includegraphics[width=0.7\linewidth, trim={10cm 5.5cm 8.5cm 5cm}, clip=True]{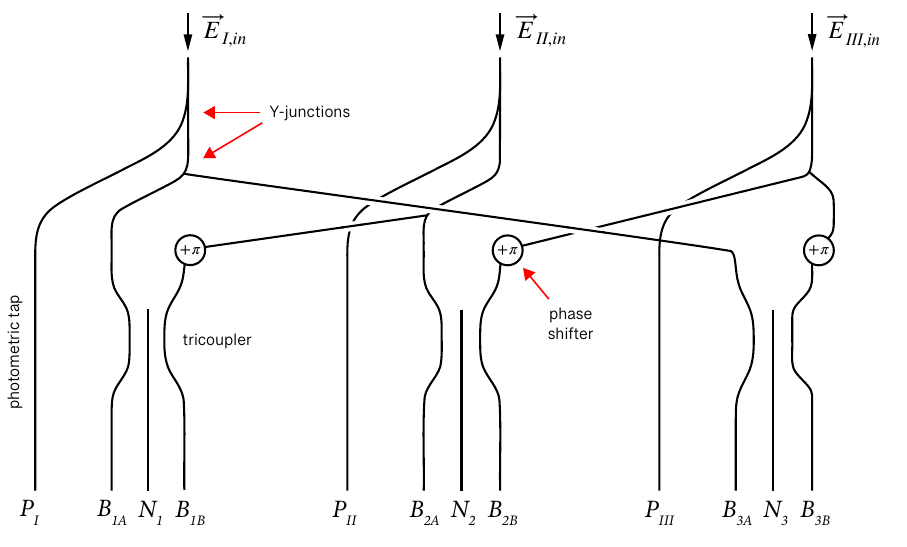}
\end{tabular}
\end{center}
\caption[example] 
{ \label{fig:chip_schema} 
A schematic of the waveguide stages in the new GLINT chip. Input fields $\vec{E}$ propagate from the top, are split at Y-junctions, and interfere at the tricoupler stage. The twelve output channels to the spectrograph are photometry channels $P$, anti-null channels across each of the formed baselines $B$, and nulled channels $N$. This is not to scale, and does not represent the waveguide side-step. Note the relative phase shift of $\pi$ is achieved by modifying the diameters of both converging waveguides.}
\end{figure}

Instantaneous phase information can be obtained with the addition of a third waveguide to each pair of coupled waveguides in a directional coupler. The third waveguide begins inside the glass bulk of the chip and threads alongside the other two waveguides in a ``tricoupler'' arrangement (Fig. \ref{fig:chip_schema}) \cite{martinod2021achromatic,klinner2022achromatic}. When the two input waveguides have a relative phase difference of $\pi$, the third waveguide will carry the nulled output, and the other two will carry off-axis light with a difference in flux that yields the phase. A fringe-tracking loop can act on this phase term and beat down the median null depth by a factor of a few tens for bright targets, according to simulations \cite{martinod2022achromatic}. An additional benefit of this wavefront sensing capability is that it is on the same path as the science channel, and would avoid non-common-path errors associated with AO configurations where the science and wavefront sensing channels split before detection. In order to enable this fringe-tracking capability, the old deformable mirror (used for static alignment and phase offsets) will be switched out for a Boston HEX-111 deformable mirror, with a PCI card interface.

In addition, the waveguides can be designed to provide a more achromatic phase shift (and by extension, the output null) by alternating their widths along different sections to change their effective index of refraction, as done by the beam combiners in VLTI/PIONIER and GRAVITY. Connecting the sections with smooth tapers minimize losses  \cite{labeye2008composants,benisty2009integrated}. (See G. Douglass et al. 13100-95 in these proceedings.) A chip with a minimum of three sub-apertures will also provide closure phases, whereby the sum of measured phases across at least three baselines cancels out systematic terms, leaving astrophysical information \cite{cornwell1989applications,monnier2000introduction}. The error on the closure phases constrains sensitivity to circumstellar structure and companions \cite{lacour2011sparse,jovanovic2012starlight}. 

The time is particularly ripe for these changes to GLINT because other improvements are being made to the upstream AO correction. SCExAO recently acquired two cryogenically-cooled HgCdTe C-RED 1 detectors. One of them will serve as the GLINT science detector, and will have a read noise of $<$1 e- compared to $<$30 e- (high gain, CDS mode) of the previous InGaAs C-RED 2 \cite{feautrier2017c,gibson2020characterization}. The Subaru facility adaptive optics system is also being upgraded with a $\sim$3000-actuator deformable mirror (DM), which will replace the 188-actuator DM in place since 2006, and will feature a new NIR wavefront sensor with a C-RED 1, and an upgrade of the current VIS curvature wavefront sensor. These changes will both improve the AO correction as well as expand the list of possible science targets, in particular redder targets such as low-mass M-dwarf stars. (See \cite{lozi2022ao3000} and Lozi et al. 13097-1 in these proceedings.)

\begin{figure} [ht]
\begin{center}
\begin{tabular}{c}
\includegraphics[width=1\linewidth]{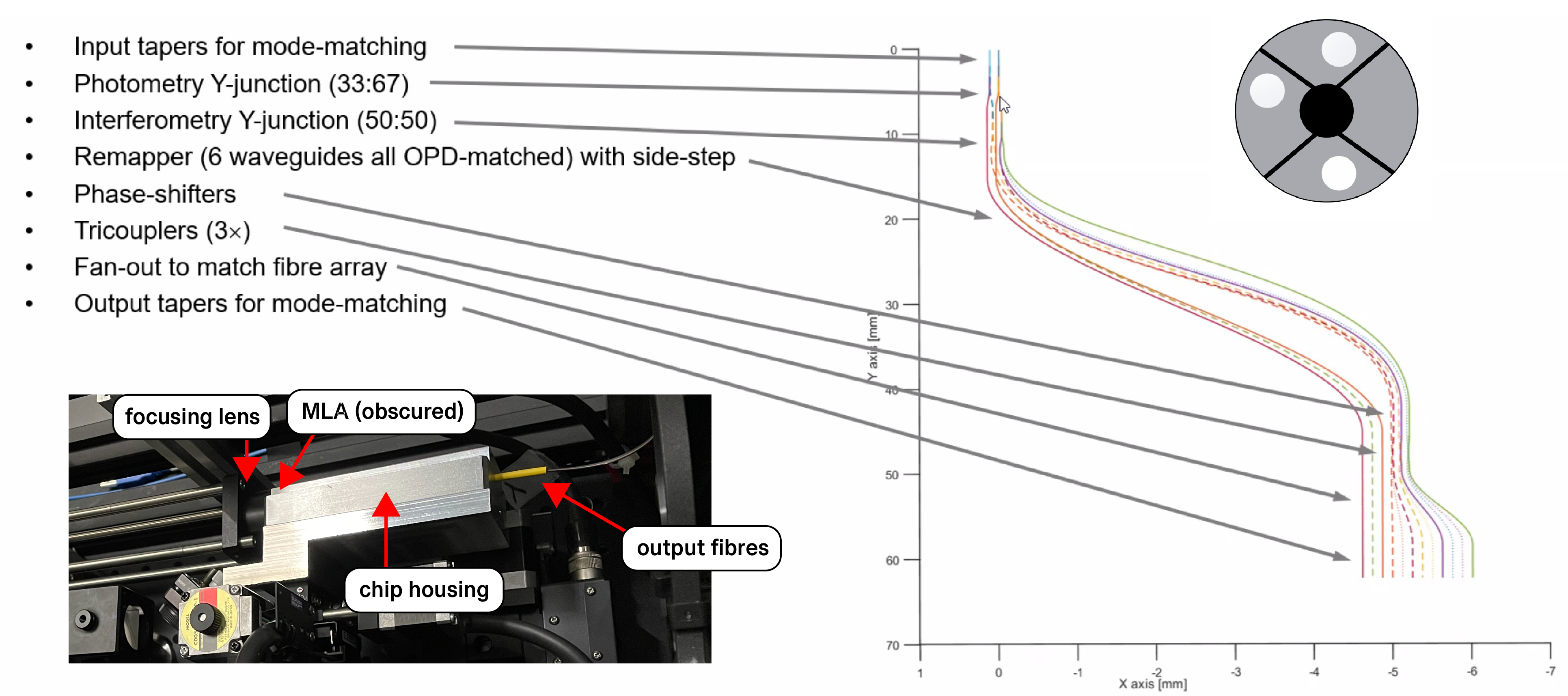}
\end{tabular}
\end{center}
\caption[example] 
{ \label{fig:chip_diag} 
A physical diagram of the waveguides with stages labeled. Note the axes have different scales. At the top, wavefronts from the three subapertures of the Subaru pupil (top right inset) propagate down and eventually exit at output fibres (seen in bottom left photo).}
\end{figure}

\section{The current upgrade}
\label{sec:title}

The new version of the GLINT chip has the primary goal of improving the null with tricouplers and achromatic phase shifters (Fig. \ref{fig:chip_schema}). A subaperture pattern of three holes was chosen to balance waveguide complexity with Fourier coverage. The center-to-center baselines are 4 to 6 meters as projected on the Subaru pupil. The holes were chosen to be circular to minimize scattered light. (The difference in throughput between circular subapertures and ones which are hexagonal, to emulate the DM segment shapes, was found to be negligible.)

For injecting the light into the chip waveguides, previous versions of GLINT used lenslet arrays which were physically offset from the optical mask and chip, and which required laborious alignment. For the new version of GLINT, the lenslets, mask, and chip are integrated into a single package:  the mask is printed onto a substrate, the lenslets are printed into the subapertures within the mask, and the glass substrate is glued to the chip. 

The lenslet substrates consisted of 0.55 mm thick non-ITO-coated Schott D263 T eco glass from UQG Optics Ltd. (Cambridge, United Kingdom). Multiple copies of optical masks were printed onto substrate slabs using black chrome with a requirement of at least optical density 4.0 at $\lambda=1.55$ $\mu$m, by Opto-Line (Wilmington, Massachusetts, USA). The dimensions of each mask were made to accommodate different permutations of deformable mirrors (Boston Micromachines Hex-111 for the instrument itself, and an IrisAO PTT111 for a testbench) and different subaperture stop sizes. The lenslets were printed using a procedure refined in-house at Macquarie University based on two-photon polymerization \cite{o2023two}, by focusing a laser into Norland 61 resin (Norland Products Inc., Jamesburg, New Jersey, USA) to form a polymer lenslet structure. 

The chip waveguides had the same basic parameters as those in \cite{norris2020first}. The first-order paths of the waveguides in the chip were designed using a nonlinear optimization, and initial ``side-step'' curves were included to minimize scattered light, as was done in previous GLINT and Dragonfly chips. The Y-splitters, phase shifters, and tricouplers were designed iteratively with a view to minimizing contaminating cross-talk, maximizing throughput, equalizing coupling, and matching pathlengths up to the tricoupler stage. Refinement was made based on test chips.

To our knowledge this marked the first time phase shifters have been made inside glass using ULI (other efforts have used lithographic techniques). After making an initial phase shifter design, waveguides of different diameters were inscribed into a test chip, and those waveguides had sets of Bragg gratings with different grating spacing (pitches) written into them. The effective indices of refraction for each diameter were measured from the reflected wavelengths of an injected source, and a polynomial was fit to the wavelength-diameter-index space. A final phase shifter design was then made based on this polynomial fit. (See Douglass et al. 13100-95 in these proceedings.)

Tricouplers were written into multiple test blocks to produce chips with different parameters such as coupling region lengths, ULI feed rates, and depths of the central waveguide. Based on the measured splitting ratios of injected light, coupled-mode equations were used to solve for coupling coefficients and a de-phasing term. The closest match to splitting coefficients which are equal, and a dephasing term which is zero, was chosen for the tricoupler design. (See Arcadi et al. 13100-247 and 13100-94 in these proceedings.) 

\begin{figure} [ht]
\begin{center}
\begin{tabular}{c} 
\includegraphics[width=1\linewidth, trim={0cm 0cm 0cm 0cm}, clip=True]{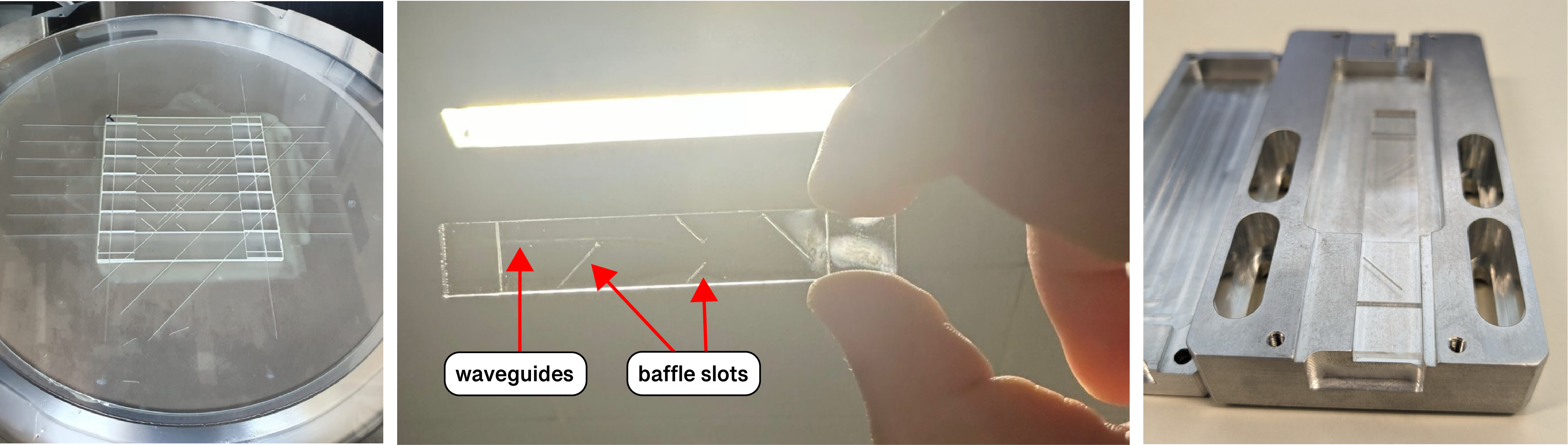}
\end{tabular}
\end{center}
\caption[example] 
{ \label{fig:chip_stages} 
Left to right: Chip copies as laser-inscribed in a single slab; a diced chip; a chip placed in the open housing.}
\end{figure}

The chip itself was made from Schott AF45 glass, with diced chip dimensions 60 mm $\times$ 6 mm. The waveguides were inscribed between about 200 and 400 $\mu$m depth by a 5.1 MHz pulsed, 800 nm Ti:sapphire laser, focused with a NA 1.25 objective lens. The chip was translated at 100 cm/min, with small variations for finer changes, and while dumping power to change the waveguide widths and induce the relative $\pi$ phase shift between the two input waveguides. To further reduce scattered light, opaque slots were inserted at diagonal angles into the chip bulk. The waveguides snake between them on their way to the output. Fig. \ref{fig:chip_diag} shows a diagram of the final chip waveguides, and Fig. \ref{fig:chip_stages} shows the physical chip.

\begin{figure} [ht]
\begin{center}
\begin{tabular}{c} 
\includegraphics[width=1\linewidth, trim={2.5cm 2.8cm 1cm 2.7cm}, clip=True]{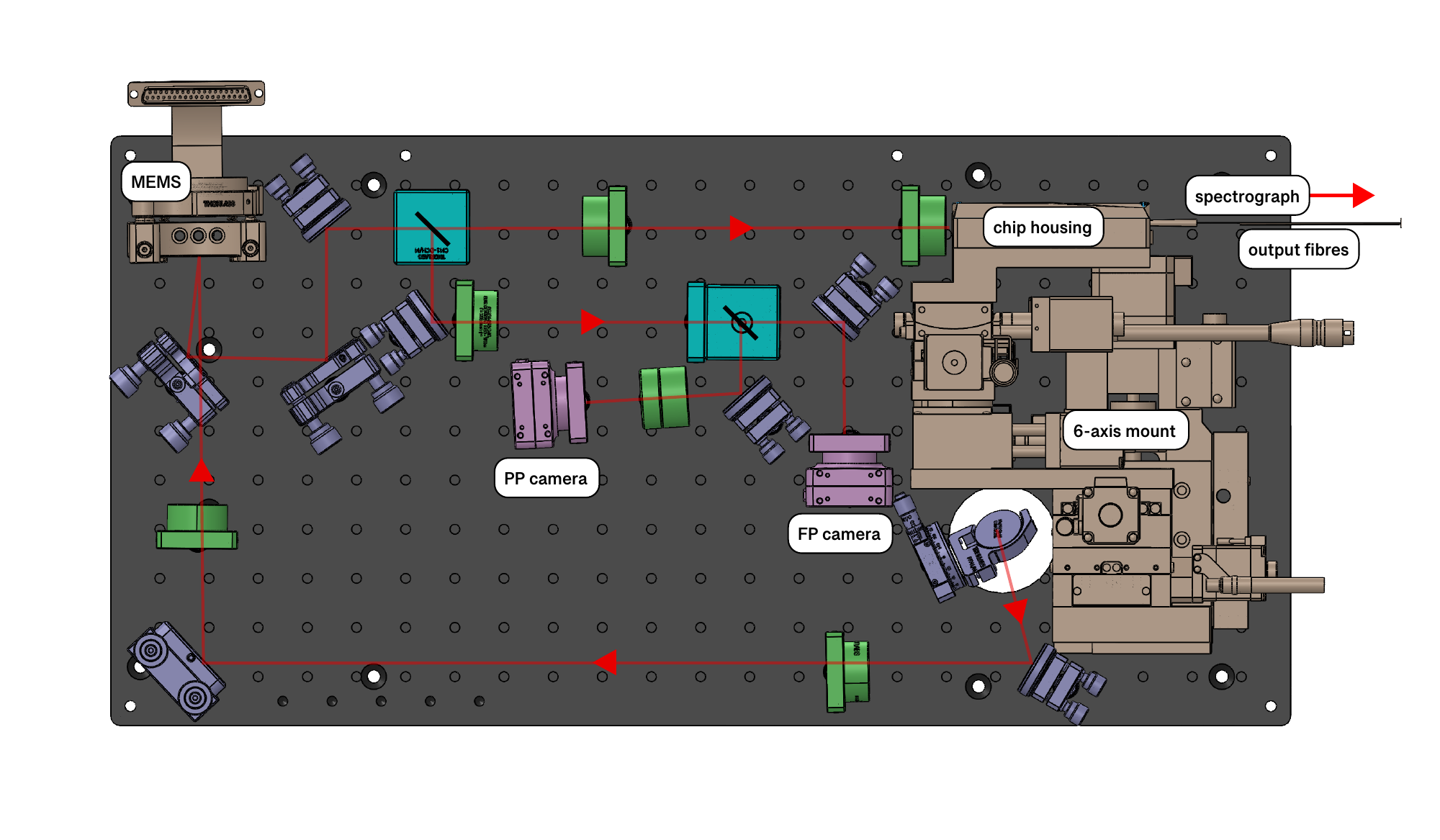}
\end{tabular}
\end{center}
\caption[example] 
{ \label{fig:layout_global} 
A CAD drawing of the current layout of GLINT}
\end{figure}

The ancillary optics of GLINT were chosen to project an undersized telescope pupil onto the DM, and a 10:1 downscaling of the pupil from the DM to the chip lenslets. This choice of downscaling accommodates limitations to the laser waveguide write depth while still keeping non-interfering sections of waveguides as far apart as possible inside the chip, except along sections where the fields are supposed to interfere. In order to have closed-loop control of the chip's movement, we designed a mount made of translation stages and goniometers from Suruga Seiki (Tokyo, Japan), fixed together with in-house-designed metal connectors cut out of aluminum by Hubs (Amsterdam, the Netherlands). This mount replaces one which only had open-loop control. (See Fig. \ref{fig:layout_global}.)

In January 2024, GLINT was upgraded at a Subaru lab in Hilo, Hawaii, with the new optics, the BM HEX-111 deformable mirror, and an initial 3-waveguide chip. Before reinstalling GLINT on SCExAO, two mirror mounts within the IR bench of SCExAO were replaced with Newport Conex stages for increased repeatability of the alignment of the beam as it exits the SCExAO IR bench and enters GLINT. After GLINT was reinstalled, the outputs of the chip were connected via a V-groove to a port that can accommodate either a C-RED 1 or C-RED 2 detector for dispersed imaging. Residual alignments using the DM segments are currently being performed with a mixture of in-person and remote work.

\section{Future plans}

Based on the lessons learned from the newly-installed 3-waveguide chip, the chip design was optimized to improve the throughput, the light splitting ratios, and the phase shifters. The lenslet manufacturing process was also refined. An optimized 3-waveguide chip was manufactured in June 2024, and will be installed at Subaru very soon. Engineering observations anticipated for later in the year will allow us to obtain on-sky performance metrics.

In the meantime, software is being developed to improve the useability of GLINT, with a GUI for control and data-taking, for both initial  observations and a science survey, and in preparation for eventually offering GLINT for open use to the Subaru community (Rossini-Bryson et al. 13095-70 in these proceedings). Another near-term software task will be to implement a fringe tracking loop based on the tricoupler outputs. 

Future chip architectures will feature more waveguides to fill in more of the Fourier coverage of the Subaru pupil\cite{tuthill2022nulling}, and a 9-waveguide chip is planned to succeed the optimized 3-waveguide chip. Two testbenches are currently being set up for characterization of chips as they are produced, and to inform future designs. One testbench at the University of Texas at San Antonio will investigate thermal phase shifting techniques. This will involve heating elements implanted in the chip in proximity to the waveguides to allow tuning of the local index of refraction \cite{ceccarelli2019thermal}. Thermal phase shifting will offload some of the phase shifting now done by the DM, and, as part of a control law, could make short-timescale perturbations to the pathlengths to deepen the null. Our objective is to make a first prototype thermally-enabled chip by the end of the year.

The other testbench is being built at the University of Sydney. It generally reflects the optical setup of GLINT at SCExAO to enable testing of alignment algorithms while freeing up use of the SCExAO bench at Subaru. This testbench will also be a platform for characterizing chips in terms of polarization effects, pathlength matching, and injection efficiencies. We also plan to simulate science targets and atmospheric seeing, so as to test fringe tracking methods.

In addition to the AO3k upgrade to the Subaru facility AO system, there are plans to install a beamswitcher on the Nasmyth platform, which will allow rapid switching of the telescope beam between SCExAO instruments without needing to crane them in and out. After the beamswitcher installation, both GLINT and AO3k's NIR WFS will be able to use C-RED 1 detectors at the same time. (Before then, GLINT will use the facility AO's visible WFS.) In the future, the modularity of the SCExAO bench could also enable the direction of the GLINT nulled channel to other detectors, such as no-noise MKIDS detectors \cite{steiger2022probing}. 

GLINT was and is a narrowband $H$-band instrument, but as the GLINT null improves, there will be increased need to consider chip architectures to broaden the bandwidth for more signal-to-noise, and to diversify to other wavelengths: either shorter, VIS/NIR wavelengths to target Hydrogen lines from accreting material in nascent planetary systems (such as in $J$-band to target the Paschen-$\beta$ line); or longer, thermal-infrared wavelengths where planet-to-star contrasts are more favorable. Further add-ons to chip architectures will become important as different systematics emerge as the next bottlenecks to the error budget. For example, fibre Bragg gratings (FBGs) can be integrated into a chip and remove narrow airglow lines, which will become important sources of atmospheric emission at wider bandpasses and at $Y$- through $K$-bands.\cite{thomson2011ultrafast,horton2012praxis,trinh2013gnosis,gross2015ultrafast,roth2016measurements,hart2018long}

Science applications for nullers on baselines across 8-meter apertures are still far from being exhausted, and nullers will have a place even before the era of ELTs. However, we can begin to take steps to extend the flexible and modular architecture of GLINT to on-chip nulling with multisegmented apertures, as prototypes for the multisegmented space-based apertures of the future; or to longer baselines, such as at the Large Binocular Telescope, which at a 22.7-meter edge-to-edge baseline is the largest such OIR baseline on a single mount (Asnodkar 13095-91 in these proceedings).

A long-baseline nuller between separate telescopes is already being planned for the VLTI. The planned Asgard instrument suite will include NOTT, an $L'$-band nuller which uses a chip with directional couplers, with a target raw null depth of $\leq$10$^{-3}$ and science objectives including the spectral characterization of exoplanets, the detection of new exoplanets in young moving groups, the investigation of exozodiacal dust, and resolving young stellar objects and AGNs \cite{defrere2018hi,defrere_review, laugier2023asgard,garreau2024asgard}. Asgard will also include Seidr, a $K$-band nuller which is being built into the Heimdallr instrument (Garreau et al. 13095-25 and Taras et al. 13095-30 in these proceedings).

Still further on in time is the space-based nulling platform Large Interferometer For Exoplanets (LIFE) (e.g., Birbacher 13095-115 and Ranganathan 13095-52 in these proceedings), which could detect a number of nearby habitable worlds \cite{quanz2022large}. The technology development we do today will inform the designs of nullers of tomorrow, and integrated photonics promises to maximize the search space for life elsewhere. Exciting times are ahead for on-chip nulling.

\acknowledgments

The development of SCExAO is supported by the Japan Society for the Promotion of Science (Grant-in-Aid for Research \#23340051, \#26220704, \#23103002, \#19H00703, \#19H00695 and \#21H04998), the Subaru Telescope, the National Astronomical Observatory of Japan, the Astrobiology Center of the National Institutes of Natural Sciences, Japan, the Mt Cuba Foundation and the Heising-Simons Foundation. 

We acknowledge support from the Australian Research Council (ARC) Discovery Program, and Astralis---Australia’s optical astronomy instrumentation Consortium---through
the Australian Government’s National Collaborative Research Infrastructure Strategy (NCRIS) Program.

The authors also wish to recognize and acknowledge the very significant cultural role and reverence that the summit of Mauna Kea has always had within the Native Hawaiian community. We are most fortunate to have the opportunity to conduct observations from this mountain.

\bibliography{report}

\begin{thebibliography}{10}

\bibitem{guerri2011apodized}
Guerri, G., Daban, J.-B., Robbe-Dubois, S., Douet, R., Abe, L., Baudrand, J., Carbillet, M., Boccaletti, A., Bendjoya, P., Gouvret, C., et~al., ``Apodized lyot coronagraph for sphere/vlt: Ii. laboratory tests and performance,'' {\em Experimental Astronomy}~{\bf 30}(1),  59--81 (2011).

\bibitem{nguyen2022gpi}
Nguyen, M.~N., Nickson, B.~F., Por, E.~H., Soummer, R., Hagopian, J.~G., Macintosh, B., Chilcote, J., Pueyo, L., Perrin, M., and Konopacky, Q., ``Gpi 2.0: optical designs for the upgrade of the gemini planet imager coronagraphic system,'' in [{\em Advances in Optical and Mechanical Technologies for Telescopes and Instrumentation V}{\nolinebreak\hspace{0.1em}]},   {\bf 12188},  1447--1458, SPIE (2022).

\bibitem{cheetham2016sparse}
Cheetham, A.~C., Girard, J., Lacour, S., Schworer, G., Haubois, X., and Beuzit, J.-L., ``Sparse aperture masking with sphere,'' in [{\em Optical and Infrared Interferometry and Imaging V}{\nolinebreak\hspace{0.1em}]},   {\bf 9907},  711--715, SPIE (2016).

\bibitem{ray2023textit}
Ray, S., Sallum, S., Hinkley, S., Sivamarakrishnan, A., Cooper, R., Kammerer, J., Greebaum, A.~Z., Thatte, D., Lazzoni, C., Tokovinin, A., et~al., ``The jwst early release science program for direct observations of exoplanetary systems iii: Aperture masking interferometric observations of the star hip 65426 at 3.8 um,'' {\em arXiv preprint arXiv:2310.11508}  (2023).

\bibitem{bland2009astrophotonics}
Bland-Hawthorn, J. and Kern, P., ``Astrophotonics: a new era for astronomical instruments,'' {\em Optics express}~{\bf 17}(3),  1880--1884 (2009).

\bibitem{gross2015ultrafast}
Gross, S. and Withford, M., ``Ultrafast-laser-inscribed 3d integrated photonics: challenges and emerging applications,'' {\em Nanophotonics}~{\bf 4}(3),  332--352 (2015).

\bibitem{jovanovic20232023}
Jovanovic, N., Gatkine, P., Anugu, N., Amezcua-Correa, R., Thakur, R.~B., Beichman, C., Bender, C.~F., Berger, J.-P., Bigioli, A., Bland-Hawthorn, J., et~al., ``2023 astrophotonics roadmap: pathways to realizing multi-functional integrated astrophotonic instruments,'' {\em Journal of Physics: Photonics}~{\bf 5}(4),  042501 (2023).

\bibitem{defrere_review}
{Defr{\`e}re}, D., {Absil}, O., {Berger}, J.~P., {Danchi}, W.~C., {Dandumont}, C., {Eisenhauer}, F., {Ertel}, S., {Gardner}, T., {Glauser}, A., {Hinz}, P., {Ireland}, M., {Kammerer}, J., {Kraus}, S., {Labadie}, L., {Lacour}, S., {Laugier}, R., {Loicq}, J., {Martin}, G., {Martinache}, F., {Martinod}, M.~A., {Mennesson}, B., {Monnier}, J., {Norris}, B., {Nowak}, M., {Pott}, J.~U., {Quanz}, S.~P., {Serabyn}, E., {Stone}, J., {Tuthill}, P., and {Woillez}, J., ``{Review and scientific prospects of high-contrast optical stellar interferometry},'' in [{\em Optical and Infrared Interferometry and Imaging VII}{\nolinebreak\hspace{0.1em}]},  {Tuthill}, P.~G., {M{\'e}rand}, A., and {Sallum}, S., eds., {\em Society of Photo-Optical Instrumentation Engineers (SPIE) Conference Series} {\bf 11446},  114461J (Dec. 2020).

\bibitem{blind2011incisive}
Blind, N., Boffin, H., Berger, J.-P., Le~Bouquin, J.-B., M{\'e}rand, A., Lazareff, B., and Zins, G., ``An incisive look at the symbiotic star ss leporis-milli-arcsecond imaging with pionier/vlti,'' {\em Astronomy \& Astrophysics}~{\bf 536},  A55 (2011).

\bibitem{persi2019infrared}
Persi, P. and Tapia, M., ``Infrared imaging of high-mass young stellar objects: evidence of multiple shocks and of a new protostar/star eclipsing system,'' {\em Monthly Notices of the Royal Astronomical Society}~{\bf 485}(1),  784--795 (2019).

\bibitem{crowe2024near}
Crowe, S., Fedriani, R., Tan, J., Whittle, M., Zhang, Y., o~Garatti, A.~C., Farias, J., Gautam, A., Telkamp, Z., Rothberg, B., et~al., ``Near-infrared observations of outflows and young stellar objects in the massive star-forming region afgl 5180,'' {\em Astronomy \& Astrophysics}~{\bf 682},  A2 (2024).

\bibitem{lykou2015dissecting}
Lykou, F., Klotz, D., Paladini, C., Hron, J., Zijlstra, A., Kluska, J., Norris, B., Tuthill, P., Ramstedt, S., Lagadec, E., et~al., ``Dissecting the agb star l2 puppis: a torus in the making,'' {\em Astronomy \& Astrophysics}~{\bf 576},  A46 (2015).

\bibitem{garufi2024sphere}
Garufi, A., Ginski, C., Van~Holstein, R., Benisty, M., Manara, C., P{\'e}rez, S., Pinilla, P., Ribas, {\'A}., Weber, P., Williams, J., et~al., ``The sphere view of the taurus star-forming region-the full census of planet-forming disks with gto and destinys programs,'' {\em Astronomy \& Astrophysics}~{\bf 685},  A53 (2024).

\bibitem{hughes2018debris}
Hughes, A.~M., Duch{\^e}ne, G., and Matthews, B.~C., ``Debris disks: structure, composition, and variability,'' {\em Annual Review of Astronomy and Astrophysics}~{\bf 56},  541--591 (2018).

\bibitem{benisty_2023}
{Benisty}, M., {Dominik}, C., {Follette}, K., {Garufi}, A., {Ginski}, C., {Hashimoto}, J., {Keppler}, M., {Kley}, W., and {Monnier}, J., ``{Optical and Near-infrared View of Planet-forming Disks and Protoplanets},'' in [{\em Protostars and Planets VII}{\nolinebreak\hspace{0.1em}]},  {Inutsuka}, S., {Aikawa}, Y., {Muto}, T., {Tomida}, K., and {Tamura}, M., eds., {\em Astronomical Society of the Pacific Conference Series} {\bf 534},  605 (July 2023).

\bibitem{muller2021water}
M{\"u}ller, J., Savvidou, S., and Bitsch, B., ``The water-ice line as a birthplace of planets: implications of a species-dependent dust fragmentation threshold,'' {\em Astronomy \& Astrophysics}~{\bf 650},  A185 (2021).

\bibitem{roccatagliata20203d}
Roccatagliata, V., Franciosini, E., Sacco, G.~G., Randich, S., and Sicilia-Aguilar, A., ``A 3d view of the taurus star-forming region by gaia and herschel-multiple populations related to the filamentary molecular cloud,'' {\em Astronomy \& Astrophysics}~{\bf 638},  A85 (2020).

\bibitem{dong2019observational}
Dong, R., Liu, S.-Y., and Fung, J., ``Observational signatures of planets in protoplanetary disks: planet-induced line broadening in gaps,'' {\em The Astrophysical Journal}~{\bf 870}(2),  72 (2019).

\bibitem{sallum2015accreting}
Sallum, S., Follette, K., Eisner, J., Close, L., Hinz, P., Kratter, K., Males, J., Skemer, A., Macintosh, B., Tuthill, P., et~al., ``Accreting protoplanets in the lkca 15 transition disk,'' {\em Nature}~{\bf 527}(7578),  342--344 (2015).

\bibitem{haffert2019two}
Haffert, S., Bohn, A., De~Boer, J., Snellen, I., Brinchmann, J., Girard, J., Keller, C., and Bacon, R., ``Two accreting protoplanets around the young star pds 70,'' {\em Nature Astronomy}~{\bf 3}(8),  749--754 (2019).

\bibitem{benisty2015asymmetric}
Benisty, M., Juh{\'a}sz, A., Boccaletti, A., Avenhaus, H., Milli, J., Thalmann, C., Dominik, C., Pinilla, P., Buenzli, E., Pohl, A., et~al., ``Asymmetric features in the protoplanetary disk mwc 758,'' {\em Astronomy \& Astrophysics}~{\bf 578},  L6 (2015).

\bibitem{currie2014deep}
Currie, T., Burrows, A., Girard, J.~H., Cloutier, R., Fukagawa, M., Sorahana, S., Kuchner, M., Kenyon, S.~J., Madhusudhan, N., Itoh, Y., et~al., ``Deep thermal infrared imaging of hr 8799 bcde: New atmospheric constraints and limits on a fifth planet,'' {\em The Astrophysical Journal}~{\bf 795}(2),  133 (2014).
\newblock doi:10.1088/0004-637X/795/2/133.

\bibitem{perrin2015polarimetry}
Perrin, M.~D., Duchene, G., Millar-Blanchaer, M., Fitzgerald, M.~P., Graham, J.~R., Wiktorowicz, S.~J., Kalas, P.~G., Macintosh, B., Bauman, B., Cardwell, A., et~al., ``Polarimetry with the gemini planet imager: methods, performance at first light, and the circumstellar ring around hr 4796a,'' {\em The Astrophysical Journal}~{\bf 799}(2),  182 (2015).
\newblock doi:10.1088/0004-637X/799/2/182.

\bibitem{weinberger2015target}
Weinberger, A.~J., Bryden, G., Kennedy, G.~M., Roberge, A., Defrere, D., Hinz, P.~M., Millan-Gabet, R., Rieke, G., Bailey, V.~P., Danchi, W.~C., et~al., ``Target selection for the lbti exozodi key science program,'' {\em The Astrophysical Journal Supplement Series}~{\bf 216}(2),  24 (2015).

\bibitem{ertel2020hosts}
Ertel, S., Defr{\`e}re, D., Hinz, P., Mennesson, B., Kennedy, G.~M., Danchi, W.~C., Gelino, C., Hill, J.~M., Hoffmann, W.~F., Mazoyer, J., et~al., ``The hosts survey for exozodiacal dust: observational results from the complete survey,'' {\em The Astronomical Journal}~{\bf 159}(4),  177 (2020).

\bibitem{turnbull2012search}
Turnbull, M.~C., Glassman, T., Roberge, A., Cash, W., Noecker, C., Lo, A., Mason, B., Oakley, P., and Bally, J., ``The search for habitable worlds. 1. the viability of a starshade mission,'' {\em Publications of the Astronomical Society of the Pacific}~{\bf 124}(915),  418 (2012).

\bibitem{golovin2023fifth}
Golovin, A., Reffert, S., Just, A., Jordan, S., Vani, A., and Jahrei{\ss}, H., ``The fifth catalogue of nearby stars (cns5),'' {\em Astronomy \& Astrophysics}~{\bf 670},  A19 (2023).

\bibitem{madhusudhan2019exoplanetary}
Madhusudhan, N., ``Exoplanetary atmospheres: Key insights, challenges, and prospects,'' {\em Annual Review of Astronomy and Astrophysics}~{\bf 57},  617--663 (2019).

\bibitem{chauvin2023direct}
Chauvin, G., ``Direct imaging of exoplanets: Legacy and prospects,'' {\em Comptes Rendus. Physique}~{\bf 24}(S2),  1--22 (2023).

\bibitem{traub2010direct}
Traub, W.~A. and Oppenheimer, B.~R.,  [{\em Direct imaging of exoplanets}{\nolinebreak\hspace{0.1em}]}, University of Arizona Press Tucson (2010).

\bibitem{bonnefoy2011high}
Bonnefoy, M., Lagrange, A.-M., Boccaletti, A., Chauvin, G., Apai, D., Allard, F., Ehrenreich, D., Girard, J., Mouillet, D., Rouan, D., et~al., ``High angular resolution detection of $\beta$ pictoris b at 2.18 $\mu$m,'' {\em Astronomy \& Astrophysics}~{\bf 528},  L15 (2011).

\bibitem{lovis2017atmospheric}
Lovis, C., Snellen, I., Mouillet, D., Pepe, F., Wildi, F., Astudillo-Defru, N., Beuzit, J.-L., Bonfils, X., Cheetham, A., Conod, U., et~al., ``Atmospheric characterization of proxima b by coupling the sphere high-contrast imager to the espresso spectrograph,'' {\em Astronomy \& Astrophysics}~{\bf 599},  A16 (2017).

\bibitem{carrion2021catalogue}
Carri{\'o}n-Gonz{\'a}lez, {\'O}., Mu{\~n}oz, A.~G., Santos, N., Cabrera, J., Csizmadia, S., and Rauer, H., ``Catalogue of exoplanets accessible in reflected starlight to the nancy grace roman space telescope-population study and prospects for phase-curve measurements,'' {\em Astronomy \& Astrophysics}~{\bf 651},  A7 (2021).

\bibitem{mamajek2024nasa}
Mamajek, E. and Stapelfeldt, K., ``Nasa exoplanet exploration program (exep) mission star list for the habitable worlds observatory (2023),'' {\em arXiv preprint arXiv:2402.12414}  (2024).

\bibitem{tuchow2024hpic}
Tuchow, N.~W., Stark, C.~C., and Mamajek, E., ``Hpic: The habitable worlds observatory preliminary input catalog,'' {\em The Astronomical Journal}~{\bf 167}(3),  139 (2024).

\bibitem{berger2001integrated}
Berger, J., Haguenauer, P., Kern, P., Perraut, K., Malbet, F., Schanen, I., Severi, M., Millan-Gabet, R., and Traub, W., ``Integrated optics for astronomical interferometry-iv. first measurements of stars,'' {\em Astronomy \& Astrophysics}~{\bf 376}(3),  L31--L34 (2001).

\bibitem{berger2003integrated}
Berger, J.-P., Haguenauer, P., Kern, P.~Y., Rousselet-Perraut, K., Malbet, F., Gluck, S., Lagny, L., Schanen-Duport, I., Laurent, E., Delboulbe, A., et~al., ``Integrated-optics 3-way beam combiner for iota,'' in [{\em Interferometry for Optical Astronomy II}{\nolinebreak\hspace{0.1em}]},   {\bf 4838},  1099--1106, SPIE (2003).

\bibitem{monnier2004first}
Monnier, J.~D., Traub, W., Schloerb, F., Millan-Gabet, R., Berger, J.-P., Pedretti, E., Carleton, N., Kraus, S., Lacasse, M., Brewer, M., et~al., ``First results with the iota3 imaging interferometer: the spectroscopic binaries $\lambda$ virginis and wr 140,'' {\em The astrophysical journal}~{\bf 602}(1),  L57 (2004).

\bibitem{lebouquin2006integrated}
Lebouquin, J.-B., Labeye, P., Malbet, F., Jocou, L., Zabihian, F., Rousselet-Perraut, K., Berger, J.-P., Delboulbe, A., Kern, P., Glindemann, A., et~al., ``Integrated optics for astronomical interferometry-vi. coupling the light of the vlti in k band,'' {\em Astronomy \& Astrophysics}~{\bf 450}(3),  1259--1264 (2006).

\bibitem{benisty2009integrated}
Benisty, M., Berger, J.-P., Jocou, L., Labeye, P., Malbet, F., Perraut, K., and Kern, P., ``An integrated optics beam combiner for the second generation vlti instruments,'' {\em Astronomy \& Astrophysics}~{\bf 498}(2),  601--613 (2009).

\bibitem{le2011pionier}
Le~Bouquin, J.-B., Berger, J.-P., Lazareff, B., Zins, G., Haguenauer, P., Jocou, L., Kern, P., Millan-Gabet, R., Traub, W., Absil, O., et~al., ``Pionier: a 4-telescope visitor instrument at vlti,'' {\em Astronomy \& Astrophysics}~{\bf 535},  A67 (2011).

\bibitem{jocou2014beam}
Jocou, L., Perraut, K., Moulin, T., Magnard, Y., Labeye, P., Lapras, V., Nolot, A., Perrin, G., Eisenhauer, F., Holmes, C., et~al., ``The beam combiners of gravity vlti instrument: concept, development, and performance in laboratory,'' in [{\em Optical and infrared interferometry IV}{\nolinebreak\hspace{0.1em}]},   {\bf 9146},  455--465, SPIE (2014).

\bibitem{perraut2018single}
Perraut, K., Jocou, L., Berger, J., Chabli, A., Cardin, V., Chamiot-Maitral, G., Delboulb{\'e}, A., Eisenhauer, F., Gamb{\'e}rini, Y., Gillessen, S., et~al., ``Single-mode waveguides for gravity-i. the cryogenic 4-telescope integrated optics beam combiner,'' {\em Astronomy \& Astrophysics}~{\bf 614},  A70 (2018).

\bibitem{hinz2000blinc}
Hinz, P.~M., Angel, J. R.~P., Woolf, N.~J., Hoffmann, W.~F., and McCarthy~Jr, D.~W., ``Blinc: a testbed for nulling interferometry in the thermal infrared,'' in [{\em Interferometry in Optical Astronomy}{\nolinebreak\hspace{0.1em}]},   {\bf 4006},  349--353, SPIE (2000).

\bibitem{colavita2010keck}
Colavita, M., Serabyn, E., Ragland, S., Millan-Gabet, R., and Akeson, R., ``Keck interferometer nuller instrument performance,'' in [{\em Optical and Infrared Interferometry II}{\nolinebreak\hspace{0.1em}]},   {\bf 7734},  264--274, SPIE (2010).

\bibitem{haguenauer2006deep}
Haguenauer, P. and Serabyn, E., ``Deep nulling of laser light with a single-mode-fiber beam combiner,'' {\em Applied optics}~{\bf 45}(12),  2749--2754 (2006).

\bibitem{serabyn2019nulling}
Serabyn, E., Mennesson, B., Martin, S., Liewer, K., and K{\"u}hn, J., ``Nulling at short wavelengths: theoretical performance constraints and a demonstration of faint companion detection inside the diffraction limit with a rotating-baseline interferometer,'' {\em Monthly Notices of the Royal Astronomical Society}~{\bf 489}(1),  1291--1303 (2019).

\bibitem{hinz2016}
{Hinz}, P.~M., {Defr{\`e}re}, D., {Skemer}, A., {Bailey}, V., {Stone}, J., {Spalding}, E., {Vaz}, A., {Pinna}, E., {Puglisi}, A., {Esposito}, S., {Montoya}, M., {Downey}, E., {Leisenring}, J., {Durney}, O., {Hoffmann}, W., {Hill}, J., {Millan-Gabet}, R., {Mennesson}, B., {Danchi}, W., {Morzinski}, K., {Grenz}, P., {Skrutskie}, M., and {Ertel}, S., ``{Overview of LBTI: a multipurpose facility for high spatial resolution observations},'' in [{\em Optical and Infrared Interferometry and Imaging V}{\nolinebreak\hspace{0.1em}]},  {Malbet}, F., {Creech-Eakman}, M.~J., and {Tuthill}, P.~G., eds., {\em Society of Photo-Optical Instrumentation Engineers (SPIE) Conference Series} {\bf 9907},  990704 (Aug. 2016).

\bibitem{jovanovic2012starlight}
Jovanovic, N., Tuthill, P.~G., Norris, B., Gross, S., Stewart, P., Charles, N., Lacour, S., Ams, M., Lawrence, J., Lehmann, A., et~al., ``Starlight demonstration of the dragonfly instrument: an integrated photonic pupil-remapping interferometer for high-contrast imaging,'' {\em Monthly Notices of the Royal Astronomical Society}~{\bf 427}(1),  806--815 (2012).

\bibitem{cherin1983introduction}
Cherin, A.~H.,  [{\em An introduction to optical fibers}{\nolinebreak\hspace{0.1em}]} (1983).

\bibitem{serabyn2021nulling}
Serabyn, E., ``Nulling interferometry,'' in [{\em The WSPC Handbook of Astronomical Instrumentation: Volume 3: UV, Optical \& IR Instrumentation: Part 2}{\nolinebreak\hspace{0.1em}]},   71--89, World Scientific (2021).

\bibitem{norris2020first}
Norris, B.~R., Cvetojevic, N., Lagadec, T., Jovanovic, N., Gross, S., Arriola, A., Gretzinger, T., Martinod, M.-A., Guyon, O., Lozi, J., et~al., ``First on-sky demonstration of an integrated-photonic nulling interferometer: the glint instrument,'' {\em Monthly Notices of the Royal Astronomical Society}~{\bf 491}(3),  4180--4193 (2020).

\bibitem{lagadec2018glint}
Lagadec, T., Norris, B., Gross, S., Arriola, A., Gretzinger, T., Cvetojevic, N., Lawrence, J., Withford, M., and Tuthill, P., ``Glint south: a photonic nulling interferometer pathfinder at the anglo-australian telescope for high contrast imaging of substellar companions,'' in [{\em Optical and Infrared Interferometry and Imaging VI}{\nolinebreak\hspace{0.1em}]},   {\bf 10701},  238--245, SPIE (2018).

\bibitem{lagadec2021glint}
Lagadec, T., Norris, B., Gross, S., Arriola, A., Gretzinger, T., Cvetojevic, N., Martinod, M.-A., Jovanovic, N., Withford, M., and Tuthill, P., ``The glint south testbed for nulling interferometry with photonics: Design and on-sky results at the anglo-australian telescope,'' {\em Publications of the Astronomical Society of Australia}~{\bf 38},  e036 (2021).

\bibitem{martinod2021scalable}
Martinod, M.-A., Norris, B., Tuthill, P., Lagadec, T., Jovanovic, N., Cvetojevic, N., Gross, S., Arriola, A., Gretzinger, T., Withford, M.~J., et~al., ``Scalable photonic-based nulling interferometry with the dispersed multi-baseline glint instrument,'' {\em Nature communications}~{\bf 12}(1),  2465 (2021).

\bibitem{norris2023machine}
Norris, B.~R., Martinod, M.-A., Tuthill, P., Gross, S., Cvetojevic, N., Jovanovic, N., Lagadec, T., Klinner-Teo, T., Guyon, O., Lozi, J., et~al., ``Machine-learning approach for optimal self-calibration and fringe tracking in photonic nulling interferometry,'' {\em Journal of Astronomical Telescopes, Instruments, and Systems}~{\bf 9}(4),  048005--048005 (2023).

\bibitem{hanot2011improving}
Hanot, C., Mennesson, B., Martin, S., Liewer, K., Loya, F., Mawet, D., Riaud, P., Absil, O., and Serabyn, E., ``Improving interferometric null depth measurements using statistical distributions: theory and first results with the palomar fiber nuller,'' {\em The Astrophysical Journal}~{\bf 729}(2),  110 (2011).

\bibitem{jovanovic2012progress}
Jovanovic, N., Tuthill, P.~G., Norris, B., Gross, S., Stewart, P., Charles, N., Lacour, S., Lawrence, J., Robertson, G., Fuerbach, A., et~al., ``Progress and challenges with the dragonfly instrument; an integrated photonic pupil-remapping interferometer,'' {\em Optical and Infrared Interferometry III}~{\bf 8445},  36--44 (2012).

\bibitem{martinod2021achromatic}
Martinod, M.-A., Tuthill, P., Gross, S., Norris, B., Sweeney, D., and Withford, M.~J., ``Achromatic photonic tricouplers for application in nulling interferometry,'' {\em Applied Optics}~{\bf 60}(19),  D100--D107 (2021).

\bibitem{klinner2022achromatic}
Klinner-Teo, T., Martinod, M.-A., Tuthill, P., Gross, S., Norris, B., and Leon-Saval, S., ``Achromatic design of a photonic tricoupler and phase shifter for broadband nulling interferometry,'' {\em Journal of Astronomical Telescopes, Instruments, and Systems}~{\bf 8}(4),  045001--045001 (2022).

\bibitem{martinod2022achromatic}
Martinod, M.-A., Klinner-Teo, T., Tuthill, P., Gross, S., Arcadi, E., Douglass, G., Webb, J., Norris, B., Guyon, O., Lozi, J., et~al., ``Achromatic nulling interferometry and fringe tracking with 3d-photonic tricouplers with glint,'' in [{\em Optical and Infrared Interferometry and Imaging VIII}{\nolinebreak\hspace{0.1em}]},   {\bf 12183},  214--226, SPIE (2022).

\bibitem{labeye2008composants}
Labeye, P., {\em Composants optiques int{\'e}gr{\'e}s pour l'interf{\'e}rom{\'e}trie astronomique}, PhD thesis, Institut National Polytechnique de Grenoble-INPG (2008).

\bibitem{cornwell1989applications}
Cornwell, T., ``The applications of closure phase to astronomical imaging,'' {\em Science}~{\bf 245}(4915),  263--269 (1989).

\bibitem{monnier2000introduction}
Monnier, J.~D., ``An introduction to closure phases,'' {\em Principles of long baseline stellar interferometry}~{\bf 203} (2000).

\bibitem{lacour2011sparse}
Lacour, S., Tuthill, P., Amico, P., Ireland, M., Ehrenreich, D., Huelamo, N., and Lagrange, A.-M., ``Sparse aperture masking at the vlt-i. faint companion detection limits for the two debris disk stars hd 92945 and hd 141569,'' {\em Astronomy \& Astrophysics}~{\bf 532},  A72 (2011).

\bibitem{feautrier2017c}
Feautrier, P., Gach, J.-L., Greffe, T., Clop, F., Lemarchand, S., Carmignani, T., Stadler, E., Doucour{\'e}, C., and Boutolleau, D., ``C-red one and c-red 2: Swir advanced cameras using saphira e-apd and snake ingaas detectors,'' in [{\em Image Sensing Technologies: Materials, Devices, Systems, and Applications IV}{\nolinebreak\hspace{0.1em}]},   {\bf 10209},  59--74, SPIE (2017).

\bibitem{gibson2020characterization}
Gibson, R.~K., Oppenheimer, R., Matthews, C.~T., and Vasisht, G., ``Characterization of the c-red 2: a high-frame rate near-infrared camera,'' {\em Journal of Astronomical Telescopes, Instruments, and Systems}~{\bf 6}(1),  011002--011002 (2020).

\bibitem{lozi2022ao3000}
Lozi, J., Ahn, K., Clergeon, C., Deo, V., Guyon, O., Hattori, T., Minowa, Y., Nishiyama, S., Ono, Y., and Vievard, S., ``Ao3000 at subaru: Combining for the first time a nir wfs using first light’s c-red one and alpao’s 64x64 dm,'' in [{\em Adaptive Optics Systems VIII}{\nolinebreak\hspace{0.1em}]},   {\bf 12185},  991--1004, SPIE (2022).

\bibitem{o2023two}
O'Halloran, S., Pandit, A., Heise, A., and Kellett, A., ``Two-photon polymerization: fundamentals, materials, and chemical modification strategies,'' {\em Advanced Science}~{\bf 10}(7),  2204072 (2023).

\bibitem{tuthill2022nulling}
Tuthill, P., ``Nulling interferometry: high contrast science for single large apertures,'' in [{\em Adaptive Optics Systems VIII}{\nolinebreak\hspace{0.1em}]},   {\bf 12185},  2709--2719, SPIE (2022).

\bibitem{ceccarelli2019thermal}
Ceccarelli, F., Atzeni, S., Prencipe, A., Farinaro, R., and Osellame, R., ``Thermal phase shifters for femtosecond laser written photonic integrated circuits,'' {\em Journal of Lightwave Technology}~{\bf 37}(17),  4275--4281 (2019).

\bibitem{steiger2022probing}
Steiger, S., Brandt, T.~D., Guyon, O., Swimmer, N., Walter, A.~B., Bockstiegel, C., Lozi, J., Deo, V., Vievard, S., Skaf, N., et~al., ``Probing photon statistics in adaptive optics images with scexao/mec,'' {\em The Astronomical Journal}~{\bf 164}(5),  186 (2022).

\bibitem{thomson2011ultrafast}
Thomson, R.~R., Birks, T.~A., Leon-Saval, S., Kar, A.~K., and Bland-Hawthorn, J., ``Ultrafast laser inscription of an integrated photonic lantern,'' {\em Optics express}~{\bf 19}(6),  5698--5705 (2011).

\bibitem{horton2012praxis}
Horton, A., Ellis, S., Lawrence, J., and Bland-Hawthorn, J., ``Praxis: a low background nir spectrograph for fibre bragg grating oh suppression,'' in [{\em Modern Technologies in Space-and Ground-based Telescopes and Instrumentation II}{\nolinebreak\hspace{0.1em}]},   {\bf 8450},  652--657, SPIE (2012).

\bibitem{trinh2013gnosis}
Trinh, C.~Q., Ellis, S.~C., Bland-Hawthorn, J., Lawrence, J.~S., Horton, A.~J., Leon-Saval, S.~G., Shortridge, K., Bryant, J., Case, S., Colless, M., et~al., ``Gnosis: the first instrument to use fiber bragg gratings for oh suppression,'' {\em The Astronomical Journal}~{\bf 145}(2),  51 (2013).

\bibitem{roth2016measurements}
Roth, K.~C., Smith, A., Stephens, A., and Smirnova, O., ``Measurements of airglow on maunakea at gemini observatory,'' in [{\em Observatory Operations: Strategies, Processes, and Systems VI}{\nolinebreak\hspace{0.1em}]},   {\bf 9910},  421--440, SPIE (2016).

\bibitem{hart2018long}
Hart, M., ``Long-term spectroscopic observations of the atmospheric airglow by the sloan digital sky survey,'' {\em Publications of the Astronomical Society of the Pacific}~{\bf 131}(995),  015003 (2018).

\bibitem{defrere2018hi}
Defrere, D., Ireland, M., Absil, O., Berger, J.-P., Danchi, W., Ertel, S., Gallenne, A., H{\'e}nault, F., Hinz, P., Huby, E., et~al., ``Hi-5: a potential high-contrast thermal near-infrared imager for the vlti,'' in [{\em Optical and Infrared Interferometry and Imaging VI}{\nolinebreak\hspace{0.1em}]},   {\bf 10701},  223--237, SPIE (2018).

\bibitem{laugier2023asgard}
Laugier, R., Defr{\`e}re, D., Woillez, J., Courtney-Barrer, B., Dannert, F.~A., Matter, A., Dandumont, C., Gross, S., Absil, O., Bigioli, A., et~al., ``Asgard/nott: L-band nulling interferometry at the vlti-i. simulating the expected high-contrast performance,'' {\em Astronomy \& Astrophysics}~{\bf 671},  A110 (2023).

\bibitem{garreau2024asgard}
Garreau, G., Bigioli, A., Laugier, R., Raskin, G., Morren, J., Berger, J.-P., Dandumont, C., Goldsmith, H.-D.~K., Gross, S., Ireland, M., et~al., ``Asgard/nott: L-band nulling interferometry at the vlti. ii. warm optical design and injection system,'' {\em Journal of Astronomical Telescopes, Instruments, and Systems}~{\bf 10}(1),  015002--015002 (2024).

\bibitem{quanz2022large}
Quanz, S.~P., Ottiger, M., Fontanet, E., Kammerer, J., Menti, F., Dannert, F., Gheorghe, A., Absil, O., Airapetian, V.~S., Alei, E., et~al., ``Large interferometer for exoplanets (life)-i. improved exoplanet detection yield estimates for a large mid-infrared space-interferometer mission,'' {\em Astronomy \& Astrophysics}~{\bf 664},  A21 (2022).

\end{thebibliography}
\bibliographystyle{spiebib}

\end{document}